
\input phyzzx

\overfullrule=0pt
\nopagenumbers
\line{PAR LPTHE 94-34 \hfil BROWN-HET-965}
\line{\hfil August 1994}
\vskip .75in
\title{{\bf COLLECTIVE HAMILTONIANS WITH\hfill
 KAC-MOODY ALGEBRAIC CONDITIONS}\foot{Work supported in part by the Department
of Energy under contract DE-FG02-91ER40688 - Task A}}
\vskip .75in
\author{{\bf Jean AVAN}\foot{Supported by CNRS; CNRS--NSF Exchange Programme
AI 0694; MICECO Exchange Programme AI 0693. Permanent address: LPTHE Paris 6
 (CNRS UA 280), Box 126, Tour 16, 1er \'etage,
4 place Jussieu 75252 Paris Cedex 05, France}{\bf and Antal JEVICKI}}
\centerline{{\it Department of Physics}}
\centerline{{\it Brown University, Providence, RI  02912, USA}}
\vskip 1.0in
\abstract

\noindent We describe the general framework for constructing collective--theory
Hamiltonians whose hermicity requirements imply a Kac--Moody algebra of
constraints on the
associated Jacobian.  We give explicit examples for the algebras $sl(2)_k$ and
$sl(3)_k$.  The reduction to $W_n$--constraints, relevant to $n$-matrix
models, is described for the
Jacobians.
\endpage

\pagenumbers

{{\bf\chapter{Introduction}}

Recent work on the collective field--theoretical approach to the description
of matrix models emphasized the particular form and consistency structure of
the
Hamiltonian [1], suggesting the use of this structure as
a guideline to the construction of Hamiltonians relevant to the
formulation of $n$-matrix models.
The proposed framework is the following:

 One considers a generalized invariant collective Hamiltonian, for a matrix or
vector theory,
of the form:
$$
{\cal H} = \sum_{i,j} \, \left(\Omega_{ij} (\phi ) \, {\partial\over
\partial\phi_i} + \omega_j (\phi ) \right)^{\dagger} {\partial\over \partial
\phi_j}\eqn\1
$$
Here $\Omega_{ij}(\phi )$ is a symmetric, field--dependent kernel.  This
Hamiltonian is non-Hermitian. It must be conjugated by the Jacobian
encapsulating  the change from the original dynamical variables
 to particular invariant collective variables.
The resulting Hamiltonian is  hermitian since the conjugation
re-establishes its unitarity properties [2]. Hermiticity conditions are
first of all the vanishing of the linear part of the Hamiltonian. This is
equivalent to the following differential equations for $J(\phi )$:
$$
\left\{ \Omega_{ij} (\phi ) \, {\partial\over \partial \phi_j } + \omega_i
(\phi ) \right\} J = 0\eqn\2
$$
  The symmetry of the
kernel $\Omega_{ij} (\phi )$ under interchange of indices $i,j$ is the second
requirement to get a hermitian Hamiltonian. This Hamiltonian reads:
$$
\eqalign{ \tilde{H} &= {1 \over 2}  \sum_{i,j} \, \left(
{\partial\over \partial\phi_i}
\Omega_{ij} (\phi ) \, {\partial\over \partial\phi_j} - {1 \over 2} {\partial^2
\Omega_{ij} \over \partial \phi_i \partial \phi_j} \right. \cr
&+ \left. {1 \over 4} (\omega_i + \sum_{k}{\partial
\Omega_{ik} \over \partial \phi_k }) \Omega_{ij}^{-1}
 (\omega_j + \sum_{k}{\partial
\Omega_{jk} \over \partial \phi_k }) \right) -{1 \over 4} \sum_{i} {\partial
\omega_i \over \partial \phi_i} }\eqn\twob
$$
 The set of
differential operators in $\2$ must  therefore
close an algebra in order for $J(\phi )$
to exist at least locally in the $\phi$-space.  Reciprocally, if one is given
a set of first order differential operators closing an algebra, the procedure
 $\1$--$\twob$ gives rise to an associated Hermitian Hamiltonian,
 provided the kernel
$\Omega_{ij} (\phi)$ can be made symmetric by suitable linear combinations
of the equations in $\2$.

In the case of 1--matrix models the operators $\2$ indeed
were identified with the positive--frequency part of
a Virasoro algebra and generalizations
related to higher matrix models were then considered [1]. In this context it is
of particular interest to realize this scheme or its generalizations
for either linear (Kac-Moody) or
 non--linear ($W_n$) algebras. These last ones arise as the relevant
 algebraic structures
for constraints determining the partition function
of higher matrix models and strings [3]. The existence of a relationship
between realizations of these two categories of algebras [4]
leads us to consider
here the simpler case of $sl(n)_k$ Kac-Moody algebras represented by first
order
differential operators.

We first of all consider the problem of
 constructing  ``collective" Hamiltonians associated, in the sense
defined by $\1$ and $\2$, to  $sl(n)_k$ Kac-Moody algebras.  We will
demonstrate that this  can be
successfully done.  The salient features of this construction,
involving a very specific form for the Jacobian $J$, are established in the
general case and illustrated by the examples of $sl(2)$, $sl(3)$
and $sl(4)$ Kac-Moody
algebras where these features arise clearly.

We  then discuss the problem of deriving $W_n$-algebra constraints in this
framework,
restricting ourselves to
 the construction of the Jacobian itself. We show how a Jacobian obeying
$W_n$-constraints follows
from a Jacobian obeying $sl(n)_k$ Kac-Moody constraints by integration over
certain well--chosen and suitably weighted variables. This in the present
context is the analogue of a Drinfeld-Sokolov reduction [5].  It represents a
``Schr\"odinger picture version" of the BRST mechanism described for example in
[6].  The construction is made explicit for $n=2$ and $n=3$.

In the conclusion, we  give some indications about the possible extensions of
the
construction, particularly to higher-order differential operators.

\medskip
{\bf\chapter{Hamiltonians Related to Kac-Moody Algebras $sl(n)_k$}}

The starting point of the construction in the general case is
the Wakimoto-Feigin-Frenkel representation
of the Kac-Moody algebra $sl(n)_k$ as a set of order-1 differential operators
[7]. It can be
obtained from the relevant currents in a $WZW$ model when the
group--valued field $g$ is adequately parametrized [8].  The Cartan generators
take the form:
$$
H_a = \sum_{k\in \, {\rm positive \; roots}} \, h_{ak} \, {\partial\over
\partial\gamma_k} +
\left(\partial_z {\partial\over \partial\phi_a} + \phi_a \right)\eqn\3
$$
The generators associated to respectively positive and negative
roots take the form:
$$
\eqalign{J_i^{\pm} = & \sum_{k\in\, {\rm positive \; roots}} \, O_{ik}^{\pm} \,
{\partial\over\partial\gamma_k} + \sum_{a\in \, {\rm Cartan}} \,
O_{ia}^{\pm} \, \left( \partial_z {\partial\over \partial\phi_a} + \phi_a
\right)\cr
& + O_i^{\pm} (\gamma , \partial_{\gamma}, \cdots ) }\eqn\4
$$
The notation is as follows.  The fields $\phi_a$ indexed by the Cartan
generators
 have an expansion on positive modes only:  $\phi_a (z)
= \sum_{n\geq 0} \, \phi_a^n \, z^n$ and their momenta have a conjugate
expansion on negative modes only ${\partial\over \partial\phi_a} = \sum_{n \geq
0} {\partial\over \partial \phi_a^n} \cdot z^{-n}$.  This gives rise to a mixed
field $(\partial_z {\partial\over\partial\phi_a} + \phi_a )$ realizing a $U(1)
\, KM$ algebra for each Cartan element.  The fields  $\gamma_k$
associated with the positive roots and their conjugate momenta ${\partial\over
\partial\gamma _k}$ have an expansion on both positive and
negative modes.  The coefficients $O_{ik}^{\pm}$ built an invertible matrix,
since
the number of independent generators must be equal to the
number of free parameters in the underlying Lie group; it
is known [7] that $h_{ak}, O_{ik}^{\pm}, O_{ia}^{\pm}$ and $ O_i^{\pm}$
depend solely on the
fields $\gamma_k$; generically the choice of suitable $\gamma_k$ coordinates
lead to $ O_i^- = 0$.

In order to use the set of $KM$ generators to construct a Hamiltonian on the
lines described in $\1$--$\twob$, we must select a maximal closed set of
differential generators from $\3$ and $\4$ in order to implement the
hermiticity
constraints $\2$ on the Hamiltonian.  We choose this maximal set to be $J_i^+
(z)$ and $H_a^{\geq 0} (z)$, that is, the positive modes in the $z$--expansion
of the Kac-Moody generator $H_a (z)$. Notice that the choice of $J_i^- (z)$
with $H_a^{\leq 0} (z)$, with the extra property $O_i^- (z) = 0$, would lead to
a trivial Jacobian, since $J_i^+ (z)$ and $H_a^{\leq 0}$ are pure
differentials.  It must be emphasized that since $J_i^+ (z)$ close a Borel
(triangular) subalgebra without central charge, one can at least formally
consider any invertible linear combination with $\gamma$-dependent
coefficients, of the functional differential equations $J_i^+ (z) \cdot J = 0$
as an equivalent admissible set of constraints, closing an algebra within the
vector space generated by $J_i^+ (z)$ with $\gamma$-dependent coefficients.
Such transformations will be crucial in our derivation.  Since only $H_a^{\geq
0}$ is considered, however, it will be practical to
restrict oneself to constant linear combinations of the functional
differential equations $H_a^{\geq 0} J = 0$. This indeed allows to maintain
the global functional notation for the differential operators and the fields.

The derivation runs as follows.  One starts with the linear constraints:
$$
\eqalign{&  \left\{ \sum_k \, O_{ik}^+ {\partial\over \partial\gamma_k} +
\sum_a  \, O_{ia}^+ \left( \partial_z {\partial\over \partial\phi_a} + \phi_a
\right) + O_i^+ (\gamma )\right\} J = 0\cr
&\left\{ \sum_k \left( h_{ak} \, {\partial\over \partial\gamma_k} \right) _{n
\geq 0} + \phi_a \right\} J = 0} \eqn\5
$$
{}From the invertibility of $O_{ik}^+$, it follows that there exists a unique
linear combination of the first ($i$-labeled) equations which gives a system
of the form:
$$
\left\{ \sum_k \, h_{ak} {\partial\over \partial\gamma_k} + \sum_b  q_{ab}
(\gamma) (\partial_z {\partial\over\partial\phi_b} + \phi_b) + h_a (\gamma )
\right\} J = 0 \eqn\6
$$
We now make an assumption, not yet proved in the general case but soon to be
checked for $n=2, 3$ and $4$:

\smallskip
\noindent\underbar{{\bf Assumption A.}}  {\it The ``coupling" matrix $(q_{ab}
(\gamma)
+ \delta_{ab} )$ is a $\gamma$-independent invertible matrix.}

It follows that the $a$-labeled equations are now reduced to purely algebraic
equalities.
$$
\sum_b \left\{ q_{ab} + \delta_{ab} ) \phi_b + h_a^{>0} (\gamma_k) \right\} J =
0 \eqn\7
$$
obtained from projecting $\6$ on positive modes and subtracting from $\5$.
The purely algebraic quantity in $\7$ is therefore written as $H_a^{>0}$ minus
a linear combination
of $J^+$.  From $\7$ it follows that the Jacobian solving $\5$ must be written
as:
$$
J = \prod_a \delta \left[ (q + {\bf 1} )_{ba}^{-1} \, h_a^{>0} (\gamma ) +
\phi_a \right] \, \tilde{J} (\gamma )\eqn\8
$$
This ansatz is then plugged back into $\5$.  Its consistency is established as
follows:

The action of the differential operators $\5$ on a form $\8$ creates terms in
$\delta (qh - \phi ) $ and $\delta ^{\prime} ( qh - \phi )$.  These last terms
arise uniquely from the action of $\5$ on the functionals $(q+ {\bf 1} )^{-1}
h^{>0} - \phi$ inside $\delta$.  This functional is, as follows from $\7$, a
linear combination of generators of the original algebra $\5$ with
$\gamma$-dependent coefficients of $J$ generators and constant coefficients of
$H$ generators. The action of differential operators in $\5$
on a functional is realized as the commutation of the generators
in $\5$ with this functional.  It gives again a functional
of $\gamma$, without dependence in $\phi_a$ since the coefficients of the
differentials in $\5$ do not depend on $\phi$.  Moreover this functional is
again a $\gamma$-dependent linear combination of the sole $J^+$ generators, by
 exact closure (without central charge) of the Borel + (positive modes of)
Cartan
algebra.  A
$\phi_a$-independent, purely functional linear combination of the generators
$J^+$ necessarily vanishes since the couplings $O_{ik}^+$ to derivatives of
$J_k^+$ are invertible. Hence no $\delta '$ term arises.

We have thus shown that the operators $\5$ can be consistently applied to
functionals of $\phi$ and $\gamma$ of the form $\8$.  The reduced set of
equations obeyed by the functional $\tilde{J} (\gamma )$  reads:
$$
\left\{ \sum_{k\in \,  {\rm roots}} \, O_{ik}^{\dagger} \,
{\partial\over\partial\gamma_k} + \sum_{a\in \, {\rm Cartan}}  O_{ia}^+ \,
\phi_a (\gamma ) + O_i (\gamma , \partial_{\gamma} \cdots ) \right\} \,
\tilde{J} (\gamma) = 0 \eqn\9
$$
Indeed the differential term ${\partial\over \partial\phi_a} $ acts only inside
the $\delta$-functional in $\8$ and we have seen that this action was exactly
compensated by the action of the ${\partial\over \partial\gamma}$ terms.
$\phi_a$ is replaced by its value $\phi_a (\gamma )$ extracted from $\7$ and
$\8$.  The set of operators $\9$ represents  the Borel subalgebra
$\5$ acting consistently on the subset of functionals $\8$; hence these reduced
operators also close the Borel (triangular) subalgebra of the
$sl(n)_k$ $KM$ algebra.

The problem now reduces to finding a linear combination of the operators
$\tilde{J}_k^+$ in $\9$ such that the kernel $O_{ik}^{\dagger}$ becomes
symmetric, in
order to be in the situation described in $\2$.  Such a coupling of
$\tilde{J}_k^+$ to ${\partial\over\partial\gamma_n}$ can always be achieved
since $O_{ik}^{\dagger}$ is invertible, as:
$$
H = \sum_{k,n,l} {\partial\over \partial\gamma_k} (z_1 ) O_{nl}^{\dagger} (z_2
) S_{nk} (z_2 , z_1 )
\tilde{J}_l^+ (z_2 )
\eqn\ten
$$
with $S_{k\ell}$ an arbitrary symmetric ``coupling matrix".  Particular forms
of $O_{ik}^{\dagger}$ may lead to more suitable parametrizations of $H$ as we
are now going to establish for $n=2$ and 3.
\medskip
{\bf \chapter{The Example of $sl(2)$ and $sl(3)$}}

The $sl(2)_k$ $KM$ algebra is represented [7] as:
$$
\eqalign{ J^+ & = - \gamma^2 {\partial \over \partial\gamma} +
 2 \gamma (\partial_z
{\partial\over \partial\phi} + \phi ) + k\, \partial\gamma\cr
H^0 & = - \gamma {\partial \over \partial\gamma} +
( \partial_z {\partial\over\partial\phi} +
\phi )\cr
J^-  & = {\partial \over \partial \gamma} }\eqn\eleven
$$
Choosing as closed algebra generators for the Hamiltonian $J^+$
and $(H^0)^{n>0}$, we consider the combination $( - {1\over \gamma} J^+ + H^0
)^{n>0}$, which gives the purely algebraic quantity $(- \phi - k
{\partial\gamma\over\gamma} )^{n>0}$ with coefficient 1 for $\phi$, thereby
realizing Assumption A.  Hence one sets the Jacobian to be:
$$
J (\phi , \gamma) = \delta \left( \phi + (k
{\partial\gamma\over\gamma})^{n>0} \right) \tilde{J} (\gamma )\eqn\twelve
$$
where $\tilde{J} (\gamma)$ obeys the (functional) equation:
$$
\left( - \gamma^2 {\partial \over \partial\gamma} + 2\gamma (- k
{\partial\gamma\over\gamma}
)^{n>0} + k \, \partial\gamma \right) \tilde{J} = 0 \eqn\thirteen
$$
or more transparently, dividing by $\gamma(z)$:
$$
 \left( - \gamma {\partial\over \partial\gamma} -
k {\partial\gamma\over\gamma} \right)_{n>0}  \tilde{J} = 0\;\;\;\;\;\;\;\;\;\;
\left( - \gamma {\partial\over \partial\gamma} + k {\partial\gamma\over\gamma}
\right)_{n<0}
\tilde{J} = 0  \eqn\fourteen
$$
The new equations cannot be written anymore as a single functional differential
equation, owing to the different nature of the reduction in $\9$
for $\phi_a$ and ${\partial\over
\partial\phi_a}$.  The generic Hamiltonian associated through $\1$ - $\2$ to
$\fourteen$ now
takes the form:
$$
H = -\int dz_1 dz_2 \gamma (z_1 ) \Omega (z_2 , z_1 ) {\partial\over
 \partial\gamma (z_1)}
{\partial\over \partial\gamma (z_2 ) }+ {\rm linear\,\, terms} \eqn\fifteen
$$
Symmetry of the double derivative kernel can be generically achieved by
writing:
$$
- \gamma (z_1 ) \Omega (z_2 , z_1 )  = - \gamma (z_2 ) \Omega (z_1 ,
z_2 )
\Rightarrow\quad\quad \Omega (z_1 , z_2 ) = \gamma (z_1) \, S (z_1 , z_2 )
\eqn\sixteen
$$
with $S (z_1 , z_2 )$ any invertible symmetric kernel.  $H$ becomes:
$$
\eqalign{  H  = -\int dz_1 dz_2  \Big\{ \gamma (z_1 ) \gamma (z_2) \, S (z_1 ,
z_2 ) &{\partial\over \partial\gamma (z_1 )}  {\partial\over \partial\gamma
(z_2 )}  \cr
 +  k \left( \left(
{\partial\gamma\over\gamma} (z_1)\right)_{n>0} - \left(
{\partial\gamma\over\gamma} (z_1)\right)_{n<0} \right)&  \gamma (z_2)
 \cdot S (z_1 , z_2 ) {\partial\over \partial\gamma (z_2)}
\Big\} }\eqn\seventeen $$
Note that a very simple choice for $ S(z_1 , z_2 )$ would be
$\delta (z_1 - z_2
)$.  From $\seventeen$ and $\twob$ one then writes the Hermitian
Hamiltonians. They have the generic form:
$$
\eqalign{  H  &= -\int dz_1 dz_2  \Big\{ {\partial\over \partial\gamma (z_1 )}
\gamma (z_1 ) \gamma (z_2) \,  S (z_1 ,
z_2 ) {\partial\over \partial\gamma (z_2 )}  \cr &+ {1 \over 4}
 \alpha (z_1) S(z_1 , z_2) \alpha(z_2) + {1 \over 2} {\delta \alpha
\over \delta \gamma}(z_1) S(z_1 , z_2)\gamma (z_2)} \eqn\seventeenb $$
where $\alpha (z)$ denotes the field ${\partial\gamma\over\gamma} (z)_{n>0}
- {\partial\gamma\over\gamma} (z)_{n<0}$. For $S = \delta$, the first two
terms in $\seventeenb$ reduce to the free quadratic Hamiltonian for the
field $\Gamma = e^{\gamma}$, since the sign shift in $\alpha$ does not
play any role when evaluating $\int dz \alpha ^2 (z)$. The term
${\delta \alpha
\over \delta \gamma}(z)\gamma (z) $ however adds a non-trivial
contribution, not to be expressed in terms of the original field
$\gamma$.

The case of $sl(3)$ is more interesting and more complicated.  The generators
of $sl(3)_k$ are parameterized [6, 8] as:
$$
\eqalign{ J_1^+ & = - \gamma_1^2 {\partial\over \partial\gamma_1} + \gamma_3 \,
{\partial\over
\partial\gamma_2} + i \alpha '\gamma_1 \left( \partial_z {\partial\over
\partial\phi_a} + \phi_a \right) +
(k+1) \partial\gamma_1\cr
J_2^+ & = (\gamma_1 \gamma_2 - \gamma_3 )\, {\partial\over \partial\gamma_1} -
\gamma_2^2
{\partial\over \partial\gamma_2} - \gamma_2 \gamma_3 {\partial\over
 \partial\gamma_3} + i \alpha ' \gamma_2
\left( \partial_z {\partial\over \partial\phi_b}  + \phi_b \right) + k
\partial\gamma_2\cr
J_3^+ & = \gamma_1 (\gamma_1 \gamma_2 -\gamma_3 ) {\partial\over
\partial\gamma_1} -
\gamma_2\gamma_3 {\partial\over \partial\gamma_2} - \gamma_3^2
{\partial \over \partial\gamma_3}
 + i \alpha ' ( \gamma_3 - \gamma_1 \gamma_2 ) \left( \partial_z
{\partial\over \partial\phi_a} + \phi_a \right) \cr &+ i\alpha '\gamma_3
( \partial_z {\partial\over \partial\phi_b} + \phi_b )
  - (k + 1) \partial\gamma_1\gamma_2 + k \partial\gamma_3\cr
H_a & = 2\gamma_1 {\partial\over \partial\gamma_1} - \gamma_2 {\partial
\over \partial\gamma_2} + \gamma_3
{\partial\over \partial\gamma_3} - \alpha '
\left( \phi_a + \partial_z {\partial\over
 \partial\phi_a} \right)\cr
H_b & = - \gamma_1 {\partial\over \partial\gamma_1} + 2\gamma_2
{\partial\over \partial\gamma_2} + \gamma_3
{\partial\over \partial\gamma_3} - \alpha '
\left( \phi_b + \partial_z {\partial\over
 \partial\phi_b} \right) \cr
J_1^- & = {\partial\over \partial\gamma_1} + \gamma_2 {\partial\over
\partial\gamma_3} \;\;\;\;\quad J_2^- =
{\partial\over \partial\gamma_2} \;\;\;\;\quad J_3^- = {\partial\over
 \partial\gamma_2}  \;\;\;\; \alpha ' = \sqrt{2k+6}}\eqn\eighteen
$$
Choosing as a closed set of differential operators the triangular upper
Borel subalgebra $J_i^+$
 plus the positive modes of the Cartan algebra
 $H_a^{n>0} , H_b^{n>0}$, the elimination of ${\partial\over
\partial\gamma}$ terms
following the general pattern $\6$,$\7$
is achieved by:
$$\eqalign{
\tilde{H}_a & = {\gamma_2\over \gamma_3 (\gamma_1 \gamma_2 - \gamma_3 )}
\left\{
(\gamma_3 - \gamma_1 \gamma_2 ) J_1^+ + {\gamma_1 \gamma_3\over \gamma_2} \,
J_2^+ - \gamma_1 \, J_3^+ \right\} - {1\over 3} (H_a - H_b )\cr
\tilde{H}_b & = {\gamma_2\over\gamma_3 (\gamma_1 \gamma_2 - \gamma_3 )} \left\{
(\gamma_3 - \gamma_1 \gamma_2 ) J_1^+ - {\gamma_1\gamma_3\over \gamma_2} \,
J_2^+ + (2 {\gamma_3\over \gamma_2} - \gamma_1 ) J_3^+ \right\} - (H_a + H_b )
}\eqn\nineteen
$$
and the $\phi$-terms are indeed coupled to a constant matrix, thereby
confirming
Assumption A:
$$\eqalign{
\tilde{H}_a^{n>0} & = ( \phi_a - \phi_b ) + 3 \left(
{\gamma_1\gamma_2\over\gamma_1\gamma_2 - \gamma_3 } \left\{ (k+1)
{\partial\gamma_1\over\gamma_1} + k {\partial\gamma_2\over\gamma_2} - k
{\partial\gamma_3\over\gamma_3} \right\} \right)^{n>0} \cr
\tilde{H}_b^{n>0} & = (-\phi_a - \phi_b ) - \left(
{\gamma_1\gamma_2\over\gamma_1\gamma_2 - \gamma_3 } \left\{ (k + 1)
{\partial\gamma_1\over\gamma_1} + k {\partial\gamma_2\over\gamma_2 } - k
{\partial\gamma_3\over\partial\gamma_3} \right\} + 2k {\partial\gamma_3\over
\gamma_3
} \right)^{n>0} }\eqn\twenty
$$
which allows us to consistently eliminate the $\phi$ dependence in $J$,
following $\8$:
$$
J (\phi_a , \phi_b , \gamma ) = \delta (\tilde{H}_a^{n>0} ) \delta (
\tilde{H}_b^{n>0} ) \, \tilde{J} (\gamma_1 , \gamma_2 , \gamma_3
)\eqn\twentyone
$$
One checks by an immediate computation that $J_{1,2,3}^+$ indeed annihilate
the functionals $\tilde{H}_a , \tilde{H}_b$.  The equations for the Jacobian
 are reduced to:
$$
\eqalign{ &\left\{ - \gamma_1^2 {\partial\over \partial\gamma_1} + \gamma_3
{\partial\over
\partial\gamma_2} + i\alpha ' \gamma_1 \left(\phi_a (\gamma_1 , \gamma_2 ,
\gamma_3
)\right) + (k+1) \partial\gamma_1 \right\} \tilde{J} = 0\cr
&\left\{ (\gamma_1 \gamma_2 - \gamma_3 ) {\partial\over \partial\gamma_1}
-\gamma_2^2
 {\partial\over \partial\gamma_2} - \gamma_2\gamma_3  {\partial\over
 \partial\gamma_3} + i \alpha ' \gamma_2
\left(\phi_b (\gamma_1 , \gamma_2 , \gamma_3 )\right)  + k \partial\gamma_2
\right\} \tilde{J} = 0\cr
& \left\{ \gamma_1 (\gamma_1 \gamma_2 - \gamma_3 ) {\partial\over \partial
\gamma_1}
 - \gamma_2 \gamma_3 {\partial\over \partial\gamma_2} - \gamma_3^2
{\partial\over \partial\gamma_3} + i
\alpha ' \gamma_3 \left( \phi_b (\gamma_1 , \gamma_2 , \gamma_3 )\right)
\right.\cr
& \left. + i\alpha ' (\gamma_3 - \gamma_1 \gamma_2 ) \phi_a (\gamma_1 ,
\gamma_2 , \gamma_3 ) + k \partial\gamma_3 - (k + 1) \partial\gamma_1 \gamma_2
\right\} \tilde{J} = 0 } \eqn\twentytwo
$$
where $\phi_a , \phi_b $ are expressed in terms of positive modes of the
functionals of $\gamma_1 , \gamma_2, \gamma_3$ obtained in $\twenty$.
The construction of the Hamiltonians associated to the set of closed
differential equations $\twentytwo$ follows from $\ten$ as $
H = \sum_{k,n,l} {\partial\over \partial\gamma_k} (z_1 ) O_{nl}^{\dagger} (z_2
) S_{nk} (z_2 , z_1 )
\tilde{J}_l^+ (z_2 )$
We recall that  $S$ is any symmetric invertible kernel --
the simplest example being $S_{ij} (z_1 , z_2 ) = \delta_{ij} \, \delta (z_1 -
z_2 ) $.

This construction is noticeably
simplified if one considers the equivalent linear
system to $\twentytwo$ obtained by diagonalizing the coefficient matrix
of the derivatives. One gets a set of equations closely resembling
 $\fourteen$:
$$
\eqalign{ &  \{\gamma_1 {\partial\over \partial\gamma_1} + \gamma_1 \gamma_2
\gamma_3 ^2 (\gamma_1 \gamma_2 - \gamma_3) \{ \phi_a - { <\phi_a> \over 2}
-k {\partial \gamma_2 \over \gamma_2} +{k \over 2}
 {\partial \gamma_3 \over \gamma_3}
\}\} \tilde{J} = 0 \cr
& \{\gamma_2 {\partial\over \partial\gamma_2} + \gamma_1 \gamma_2
\gamma_3 ^2 (\gamma_1 \gamma_2 - \gamma_3) \{ \phi_a -{ <\phi_a> \over 2}
-(k+1) {\partial \gamma_1 \over \gamma_1} -{k \over 2}
{\partial \gamma_3 \over \gamma_3}
\}\} \tilde{J} = 0 \cr
 & \left\{2\gamma_3 {\partial\over \partial\gamma_3} +\gamma_3 ^2 (\gamma_1
\gamma_2 - \gamma_3)(\gamma_1 \gamma_2 - 2\gamma_3)
  \{ \phi_a - { <\phi_a> \over 2}
+(\phi_b -{ <\phi_b> \over 2})\} \right.\cr
&+ \left. \gamma_1 \gamma_2
\gamma_3 ^2 (\gamma_1 \gamma_2 - \gamma_3) \{ \phi_a -{ <\phi_a> \over 2}
-(\phi_b -{ <\phi_b> \over 2})
-k {\partial \gamma_2 \over \gamma_2} +(k+1) {\partial \gamma_1 \over \gamma_1}
 \} \right\}  \tilde{J} = 0 } \eqn\twentytwob
$$
where $<\phi_{a,b}>$ denoted the expression of $\phi_{a,b}$ in terms of
the field $\gamma$ obtained from $\nineteen$ {\it without} projecting out the
negative modes:
$$
\eqalign{ & <\phi_{a}> = -2
{\gamma_1\gamma_2\over\gamma_1\gamma_2 - \gamma_3 } \left\{ (k+1)
{\partial\gamma_1\over\gamma_1} + k {\partial\gamma_2\over\gamma_2} - k
{\partial\gamma_3\over\gamma_3} \right\} -k {\partial\gamma_3\over\gamma_3} \cr
& <\phi_{b}> =
{\gamma_1\gamma_2\over\gamma_1\gamma_2 - \gamma_3 } \left\{ (k+1)
{\partial\gamma_1\over\gamma_1} + k {\partial\gamma_2\over\gamma_2} - k
{\partial\gamma_3\over\gamma_3} \right\} -k {\partial\gamma_3\over\gamma_3}}
\eqn\twentytwoc
$$
 Hence $\phi_a -1/2 <\phi_a>$ is exactly the difference
between positive and negative modes of $<\phi_a>$, a combination
which already occurred in
the case of $sl(2)$, see $\seventeen$.

The general $sl(3)$ Hamiltonian thus follows from $\twentytwoc$ and $\ten$,
and the explicit Hermitian Hamiltonian from $\twob$. It
will exhibit the same features as in the $sl(2)$ case: a free quadratic
part in the fields $\gamma$ and its conjugate
${\partial \over \partial \gamma}$,
and a non-trivial contribution from the $\gamma$ functional derivative
of the split field $<\phi_{a,b}>_{n<0} - <\phi_{a,b}>_{n>0}$.

Using the formulae given in [8] we have also proved the validity
of Assumption A for the algebra $sl(4)_k$. We do not include the explicit
proof here due to its lengthiness. It follows from long but
straightforward computations on similar lines as $\eighteen$-$\twenty$.

The construction in Section 2 is now safely established for $sl(2,3,4)$.
Moreover the non-triviality of the $sl(3,4)$ cases and the simplicity
of the actual linear combinations required to prove Assumption A
leads us to expect that it can be proved on general
grounds for all $KM$ algebra representations of the type $\3$,$\4$, thereby
validating the whole scheme at one strike.
\medskip
{\bf\chapter{From Kac-Moody to $W$-algebras.  The Jacobian}}

Once one has obtained a Jacobian associated with a set of Kac-Moody generators,
as above for $sl (2)$ and $sl(3)$, one can extract from it a ``Jacobian"
associated with a different, non-linear algebra of higher-order differential
operators which can be shown in the cases of $sl(2)$ and $sl(3)$ to realize the
$W_2$ (Virasoro) and $W_3$ algebra (more exactly  a centerless sector of these
algebras). The general procedure is as follows.

One goes back to the full Jacobian $\7$ obeying the complete set of $sl(n)_k$
$KM$-operators equations.  One then implements on this Jacobian the {\it
classical}
reduction leading from the $KM$ to a $W_n$--algebra.  Classically this is
achieved
by setting a particular subset of momenta $\beta_i ={\partial\over
 \partial\gamma_i}$ to
constant values 0 or 1, the particular choice corresponding to particular $W_n$
algebras [4].  This will correspond in our formalism to integrating out a
subset  of
variables, thereby deriving an effective Jacobian depending on the reduced set.
Specifically we
shall  integrate with respect to $\gamma$--variables all equations of the
$KM$ set, adding a pre-factor $e^{\gamma_{i}^{0}}$ each time one classical
momentum $\beta_i$ should be put to 1. In this way, the derivatives
${\partial\over\partial\gamma_i}$ are formally put to their reduced values 0 or
1.  The result
is generically a set of coupled differential equations for momenta of $J$.
The assumption we make, proved here for $sl(2)_k$ and $sl(3)_k$, is that
this set of integral differential equations can be transformed into a set of
purely differential equations for the integrated density $\int d\gamma_1 \cdots
d\gamma_n e^{\gamma_{i}^{0}} \cdots J(\phi_a \cdots  \gamma_1 \cdots
\gamma_n )$.  Since the equations one originally gets are equations relating
various momenta of this density, (i.e. $\int d\gamma_1 \cdots d\gamma_n \,
\gamma_i^{p_1} \cdots \gamma_n^{p_n} \, e^- \, J (\gamma_1 \cdots \gamma_n ,
\phi_a ,
\phi_n ))$ this means that we conjecture the existence of (rank $sl(n)$)
combinations of the differential operators $(J_i^+ , H_a^+$ ) where {\it all}
higher momenta of $J$ are eliminated.

\noindent {\bf a) \underbar{$sl(2)$ case}}

The Jacobian obeys the two sets of equations:
$$
\left( - \gamma^2 {\partial\over \partial\gamma} + 2\gamma (\phi + \partial_z
{\partial\over
\partial \phi}) + k \partial\gamma \right) J  = 0\eqn\twentyfour
$$
$$
\left( - \gamma {\partial\over \partial\gamma} + \partial \varphi
\right)_{n\geq 0} J  = 0
\eqn\twentyfive
$$
{}From $\twentyfour$, integrating over $\gamma$ after multiplying by
$e^{\gamma_{0}}$, one
gets
$$
\int e^{-\gamma_{0}} ( - \gamma^2 + 2\gamma A + 2\gamma \, \partial\varphi +
\partial\gamma ) J =
0 \eqn\twentyfive
$$
$A$ is a regularized constant, formally equal to $\sum_{-\infty}^{+\infty} 1$
and arising
from partial integration of a term $\gamma {\partial\over
 \partial\gamma}$. From now on $\partial\varphi$
will be understood as a  short notation for $(\phi +
\partial_z {\partial\over\partial\phi})$.It denotes a field of conformal spin
1.
 From $\twentyfive$ one
gets first,
$$
\int e^{-\gamma_0} ( - \gamma + A + \partial \varphi )_{n\geq 0} J =
0\eqn\twentysix
$$
and multiplying $\twentyfive$  by $(-\gamma + \partial\varphi )_p$ for $p \in
{\bf Z}$.
$$
\forall p ,\forall n \geq 0 \int e^{-\gamma_{0}} \left( ( - \gamma + \partial
\varphi
)_p (-\gamma + \partial\varphi )_n + (-\gamma + \partial \varphi )_p
\delta_{n,0} A -
\gamma_{n+p} \right) J = 0 \eqn\twentyseven
$$
Summing now $\twentyseven$ for all values of $p, n\geq$ 0 such that $n+p =
m\geq 0$ gives
$$
\int e^{-\gamma_{0}} \left( ( - \gamma + \partial\varphi )_m^2 - (A-m) \gamma_m
-A^2 \delta_{m,0} \right) J = 0\eqn\twentyeight
$$
Indeed, in order to get $(-\gamma + \partial\varphi )_m^2$ as a result of
the summation, one needs to sum for $n$ positive from ${m\over 2} + 1$ to
$\infty$   twice and $n = {m\over 2}$ once.  This explains the $(A-m)$ factor
in front of $\gamma_m$ as $\sum_{-\infty}^{+\infty} 1 (= A) - 2
\sum_{-m/2}^{m/2} 1 (= m)$.

The $m\gamma_m$ term turns into $\partial\gamma$ which will be replaced by
$\partial^2 \varphi$ using $\twentysix$ and therefore will contribute to the
final value of the central charge.

Putting together now $\twentyeight$, $\twentysix$ and $\twentyfive$ results in
a
 complete elimination of
the higher momenta of $J$, leaving only:
$$\int e^{\gamma_{0}} \left( (\partial\varphi + A/2 )^2 - A^2/4 + (k+1)
\partial^2
\varphi \right)_{m\geq 0} J = 0\eqn\twentynine
$$
from which the weighted density $J$ is immediately seen to obey a set
of Virasoro constraints $\sim [ (\partial \varphi )^2 + (k+1) \partial^2
\varphi ]^+$
without a central term.

\noindent{\bf \underbar{sl(3) case}}.

This case is more interesting and complicated.  One can however check
that there exist a suitable combination of generators eliminating the highest
momenta.  Indeed one has (setting the conjugate fields $\beta_{1,2} =
{\partial\over \partial\gamma_{1,2}}$ to 1 and $\beta_3
={\partial\over\partial\gamma_3}$ to 0 by
using the weighted integration measure $d\gamma_1 d\gamma_3 d\gamma_3
e^{\gamma_{1}^{0}}
e^{\gamma_{2}^{0}}$)
$$
\eqalign{& (2 H_a + H_b ) (2H_b + H_a) (H_a - H_b ) - (J_3 - 27 (H_a + H_b )
J_2 + 9 (H_a + 2H_b) (J_1 + J_2 ))\cr
& = (2\partial\varphi_a + \partial\varphi_b ) (2\partial\varphi_b +
\partial\varphi_a ) (\partial\varphi_a - \partial\varphi_b
) + 9 (\partial\varphi_a + 2\partial\varphi_b ) \partial\gamma_1 }\eqn\thirty
$$
It follows that one should consider, following a similar
 scheme as in $\twentyseven$, the realization
of the positive modes of this generator by taking suitable momenta of the
equations $(H_a - H_b )^{\geq 0} J = 0, (2H_a + H_b )^{\geq 0} J = 0 ,
 (2H_b + H_a )^{\geq 0} J = 0$.
 Such a realization, as we have seen in $\twentyeight$, modifies the
non-derivative terms
$k\partial\gamma_i$ by constant shifts of $k$.  Extra terms due to part
integration also arise.  The statement now is that the leading order of
 relevant momenta of the density $e^{\gamma_0} J$ is eliminated by this
suitable
combination of generators.  We conjecture that this statement   be true at
lower
orders -- the extra term has exactly the form $F(\phi_a) \partial\gamma_i$
which the part-integration and index-shift effects generate -- in which
case the integrated density is annihilated by the positive modes of a $W_3$
current:
$$
W = (2\partial\varphi_a + \partial\varphi_b ) (2\partial\varphi_b +
\partial\varphi_a )
 (\partial\varphi_a - \partial\varphi_b)   \eqn\thirtyone
$$
$\thirtyone$  is indeed the canonical realization of a $W_3$ current.

The Virasoro algebra part is obtained as follows:
The combination of generators $H_a^2 + H_b^2 + H_a H_b + 3 (J_1 + J_2 )$ yields
:
$$
L=(\partial \varphi_a )^2 + (\partial \varphi_b) ^2 + \partial \varphi_a
\partial
\varphi_b
  + 3 (k+1) \partial\gamma_1 + 3
\partial\gamma_2 \, . \eqn\thirtytwo
$$
It follows that up to lower-order terms in $\gamma$, this combination will
allow to eliminate all $\gamma$--momenta of $\int e^{\gamma_{0}} J$.  But
lower-order here means only linear, and linear momenta in $\gamma$ can be
re-expressed as linear momenta in $(\phi + \partial_z{\partial\over
 \partial\phi})_{a,b}$
from $H$-equations.  Hence the complete elimination of
$\gamma$-momenta is proved to be achieved for the Virasoro generators.

We would like to conclude this section by reminding that the Jacobian $J$ is
necessarily of the form $J = \delta (\phi - \phi (\gamma ) ) \tilde{J}
(\gamma)$.  Hence integration of $J$ with respect to $\gamma$ will yield very
non-trivial
dependences on the remaining Cartan algebra parameters $\phi_a$.

\medskip
{\bf\chapter {Conclusions and Extensions}}
Let us briefly summarize  the scheme and the construction presented in
this paper. We have described a construction of general $sl(n)_k$-based
collective
hamiltonians. To this effect  we have postulated the existence of a consistent
construction by which,
choosing a Borel subalgebra of a Kac-Moody algebra $sl(n, {\bf C})$
generated by the set $ \{J_a, H \}$
one obtains a corresponding Jacobian obeying $\2$, expressed as
 $J= \delta(\phi_a - \phi_a(\gamma)){\tilde J}
(\gamma)$. The algebra of constraints obeyed by this Jacobian was then
reformulated
in two ways:

1)  a reduced linear strictly triangular subalgebra, obtained
 by eliminating trivially the $\phi_a$
variables following $\8$ : $J = {\tilde J} (\gamma), \phi_a = \phi_a (\gamma),
{\partial\over \partial\phi_a}$ drops out.

2)  a nonlinear $W_n$ algebra (positive-index part) obtained
by eliminating the $\gamma$ variables
 by integration, leading to $J = J( \phi_a)$.

We have then explicitely demonstrated the consistency of the
scheme for $sl(2)_k$, $sl(3)_k$ and partially for  $sl(4)_k$.

One obvious generalization to be
achieved now is the extension of this proof to all $sl(n)_k$ by proving the
global
validity of our assumptions, and the subsequent construction of
 the $sl(n)_k$-Hamiltonians by the
already well-established scheme leading to $\fifteen$ and
$\twentytwo$.
This should follow from the WZW represntation of KM algebras
derived in [8].

These theories will be useful as generalized field
theories of matrix models and strings. It was seen in [1] that through
stochastic quantization matrix models lead to generalized (loop
space) Hamiltonians, the Jacobians being identified with the partition
function.
Based on the constraints obeyed by the present Jacobians and following
procedure 2) given above we see the emergence of the $W_n$ algebra of
equations constraining the partition function of $n$-matrix models.

 Finally it is relevant to state how the scheme $\1$-$\2$
can be extended to Hamiltonians involving differential operators of higher
order. The general statement reads:

For any differential operator of order $n$:
$$
H= \sum_{k=0}^{n} O_{i_1 ... i_n}^{(k)} (\gamma )  \prod _{r=1}^k  {\partial
 \over \partial \gamma_{i_r}}
$$
the even-order (resp. odd--order) terms can be eliminated and the operator
made antihermitian (resp. hermitian) for $n$ odd (resp. even) provided a
solution exists to the set of equations:
$$
\forall p = 1 ... n/2 ,
$$
$$
([\prod _{k=1}^p {\partial \over \partial \gamma_{i_k}}] O_{i_1 ...
i_{n-p+1}}^{(n-p+1)}
- [\prod _{k=1}^{p-1} {\partial \over \partial \gamma_{i_k}}] O_{i_2 ...
i_{n-p+1}}^{(n-p)}
+...+  O_{i_{p+1} ... i_{n-p+1}}^{(n-2p+1)}) J = 0 \eqn\aa
$$
 and provided the odd-order (resp. even-order) coefficients $O^{(k)}$ are
totally symmetric under permutation of their indices. The solution to this
system
is the looked-for Jacobian.

This extension should allow a direct study, at the Hamiltonian level,
of the case of $W_n$ algebras, represented as they are by algebras
of $n$-th order differential operators, to be identified from
$\aa$. This  will be adressed in a forthcoming study.

\smallskip
{\bf Acknowledgements} We wish to thank the referee for fruitful suggestions.
J. A. thanks all members of Brown University Physics Department for their kind
hospitality.
\smallskip

\noindent{\bf References}

\pointbegin
A. Jevicki and J. P. Rodrigues, {\it Nucl. Phys.} {\bf B421} (1994) 278;
\point
A. Jevicki and B. Sakita, {\it Nucl. Phys.} {\bf B165} (1980) 511; \nextline
S. Das, A. Jevicki, {\it Mod. Phys. Lett.} {\bf A 5} (1990), 639.
\point
 R. Dijkgraaf, E. Verlinde and H. Verlinde, {\it Nucl. Phys.} {\bf B348} (1991)
435;\nextline
M. Fukuma, H. Kawai and R. Nakayama, {\it Int. J. Mod. Phys.} {\bf A6} (1991)
1385;\nextline
E. Gava, K.S. Narain, {\it Phys. Lett} {\bf B 263} (1991), 213.
\point
F. A. Bais, P. Bouwknegt, M. Surridge and K. Schoutens, {\it Nucl. Phys.} {\bf
B304} (1988) 348;\nextline
J. Balog, L. Feher, P. Forgacs, L. O'Raifeartaigh, and A. Wipf, {\it Ann. Phy.}
 {\bf 203} (1990) 76.
\point
V. G. Drinfeld and V. Sokolov, {\it Sov. J. Math} {\bf 30} (1985) 1975.
\point
M. Bershadsky and H. Ooguri, {\it Comm. Math. Phys.} {\bf 126} (1988)
49;\nextline
P. Furlan, A. Ch. Ganchev and V. B. Petkova, {\it Phys. Lett.} {\bf B318}
(1993) 85.
\point
M. Wakimoto, {\it Comm. Math. Phys.} {\bf 104} (1986), 605; \nextline
B. L. Feigin, E. V. Frenkel, {\it Russ. Math. Surveys} {\bf 43} (1989), 221;
\nextline
P. Bouwknegt, J. Mc Carthy, K. Pilch; {\it Com. Math. Phys.} {\bf 139} (1990),
125.
\point
A. Gerassimov et al. {\it Intern. Journ. Mod. Phys.} {\bf A5} (1990),
2495;\nextline
A. Yu. Morozov, {\it Phys. Lett.}{\bf B229} (1990), 239.

\bye